\documentstyle[psfig,aps,epsf]{revtex}
\begin{document}
\author{\textbf{N. Caticha}$^{*}$ \textbf{, J. E. Palo Tejada}$^{*}$
\textbf{, D. Lancet}$^\diamondsuit $ and
\textbf{ E. Domany}$^{\#}$
  \\
$^{*}$Instituto de F\'{\i}sica, Universidade de S\~{a}o Paulo, CP66318, \\
CEP 05315-970, S\~{a}o Paulo, SP, Brazil\\
$^{\#}$Department of Complex Systems, \\
Weizmann Institute of Science, Rehovot 76100, Israel \\
$^\diamondsuit $Department of Membrane Research and Biophysics,\\
Weizmann Institute of Science, Rehovot 76100, Israel}
\title{Computational Capacity of an Odorant Discriminator: the Linear
Separability of Curves}
\date{August 2000 }
\maketitle

\begin{abstract}
We introduce and study 
an artificial neural network, inspired by the probabilistic
Receptor Affinity Distribution model of olfaction. Our system consists on 
$N$ sensory neurons
whose outputs converge on a single processing linear threshold element.
The system's aim is to
model discrimination of a single target odorant from a large number
$p$ of background odorants,
within a range
of odorant concentrations. We show that this is possible provided $p$ does not
exceed a critical value $p_c$, and calculate the critical capacity 
$\alpha_c = p_c/N$. The critical capacity depends on the range of concentrations
in which the discrimination is to be accomplished. If the olfactory bulb may be 
thought of as a collection of such processing elements, each responsible for the
discrimination of a single odorant, our study provides a quantitative
analysis of the
potential computational properties of the olfactory bulb.
The mathematical formulation of the problem we consider is one of 
determining the capacity for linear separability 
of continuous curves, embedded in a large dimensional
space. This is accomplished here by a numerical study, using a 
method that signals whether the
discrimination task is realizable or not, together with a finite size
scaling analysis.
\end{abstract}

\section{Introduction}
The basic machinery for
olfaction, our ability to smell, is an array of a few hundred 
different types of sensory
neurons. Each of these expresses  molecular receptors, that belong to 
a single type. When this small neuronal assembly is exposed to
external stimuli, its
cooperative response 
is capable to detect and recognize a wide variety of {\it odorants} and to
measure their concentrations. We use the terminology ``odorant" to describe
any chemically homogenous substance (ligand)
which elicits a response from the olfactory
system.

The response of the array of neurons to any particular odorant
is determined by the responses of the individual constituent neurons.
This response is, however, governed by the extent to which the receptors
expressed by the particular neuron bind the odorant, i.e. by the {\it affinity}
$K$ of the neuron's receptors to the odorant. According to a recently
proposed model~\cite{RAD}, these affinities 
can be viewed as independent random variables,
drawn from a single receptor affinity distribution (RAD), denoted by
$\psi (K)$. Once a set of affinities (for all odorants and all sensory neurons)
has been generated, the response of the entire sensory assembly to 
any odorant is  determined.

This information is transferred from the sensory neurons to the
olfactory bulb, onto which the axons of the sensory neurons project. They form
synapses on secondary neurons  (mitral and
tufted cells). This integration of the sensory input, that takes place in
the olfactory bulb, forms the
first step of the information processing 
that takes place in the olfactory pathway. 
Interneurons of two major types (periglomerular and granule
cells) are believed to play a role in computing the pattern transmitted from
the olfactory bulb to higher brain centers.

In this paper we evaluate, on the basis of a very simple model, some of
the potential computational characteristics of the olfactory bulb,
as it performs this initial integration.
We hope some of our quantitative results could be biologically relevant.
Our simple model for the sensory array and a single processing unit is depicted
in Fig. \ref{Fig1}.

The model we introduce is, however, interesting  
also from a mathematical point of view.
The problem of Linear Separability (\textbf{LS}) of
points in $N$ dimensional space has received considerable attention
since the 19th century \cite{swiss}. In the mathematics literature
Cover studied the problem of \textbf{LS} of
independent dichotomies using combinatorial methods ~\cite{Cover}. 
In computer science the perceptron, introduced by Rosenblatt~\cite{Rose}
and analyzed in detail by Minsky and Papert~\cite{Minsky}, gave a major
boost to the field of
neural networks. More recently, by introducing
Statistical Mechanics techniques Gardner~\cite{Gardner} 
extended Cover's
results to cases where there are correlations between 
the points that have to be linearly separated.

We generalize the problem of separating 
(zero-dimensional) {\it points}, 
to the
separability of (one-dimensional) {\it strings} or {\it curves},
embedded in $N$-dimensional space. 
In the context of our problem the curves that need be separated 
are parametrized continuously by the odorant concentration.

In principle one can address the separability of curves by placing a discrete
set of points on each curve, thereby mapping the problem onto the previously
solved one, of separating points. One should note,
however, that points that lie on the same curve are not independent; in fact
they are correlated in ways that render the 
previously developed analytical methods unapplicable.
Therefore we present an extensive
numerical analysis of the capacity of this special neural network, of $N$
sensory neurons that provide input to a single processing unit.
The capacity we calculate is interpreted as follows. The sensory system is 
exposed to $p+1$ odorants, {\it one at a time}. 
One of these is the ``target"; the
aim is to distinguish the target from all the other
$p=\alpha N$ odorants that form a ``noisy olfactory background".

The model, based on a single layer perceptron, is introduced and discussed
in detail in section 2.1 . Then we turn to describe 
the method we have developed
in order to determine
numerically the capacity. To do this we had to adapt and use several 
different techniques. One of these,
a learning algorithm 
introduced by
Nabutovsky and Domany~\cite{ND}, is described in Sec 2.2. 
This algorithm, like all other
perceptron learning rules, finds
the separation plane (if the problem {\it is} \textbf{LS}); however, unlike other 
learning algorithms, it provides a rigorous signal 
to the fact that a
sample of examples is {\it not} \textbf{LS}.

Another technique we had to adapt to our purposes is finite size scaling
(FSS) analysis of the data. The main
results are presented in Sec 3 as curves of capacity as a function of odorant
concentration in the thermodynamic ($N \rightarrow \infty$)
limit, obtained by extrapolation, using FSS, 
from data obtained at a sequence of $N$ values. 
This large $N$ limit is quite natural from both practical and theoretical
points of view. In practice, for $N$ of the order of a few hundred, the
results can hardly be distinguished numerically from those at the 
$N \rightarrow \infty$ limit. As to the 
theoretical side, the situation in this limit is much cleaner and easier 
to analyze. The
final section 4 contains a critical discussion of the results from a
biological point of view.

\begin{quote}
Our central finding is summarized in Fig \ref{fig:alphac}; if we fix the range
of concentrations in which the system operates, and increase the number of 
background odorants, we will reach a critical number $p_c$ beyond which the 
system fails to discriminate the target. This critical number is proportional
to the number of sensory neurons $N$, i.e. $p_c=\alpha_c N$, and it 
decreases when the concentration range increases.
\end{quote}

\section{Computational Model}

\subsection{Odorant identification as Linear Separation of Curves in N-dimensional space}
The simple neural assembly that is considered here
consists of a single secondary neuron, which
receives inputs from an array of $N$ units that model the sensory neurons.
The single secondary neuron represents a ``grandmother cell", whose task is to
detect one
particular ``target'' odorant, labeled 0. 
The sensory scenario  we consider allows exposure of the neuronal assembly to a
single odorant, which may either be the target odorant or one of 
$p=\alpha N$ background odorants. The odorant 
provides simultaneous stimuli to the $N$ sensory
neurons. The aim of the single secondary neuron is to determine whether the
odorant that generated the incoming signal from the sensory array is  
the target odorant 0 or not. 
We assume that all odorants, background and target, 
are presented to the sensory array 
in concentrations $H$ that lie within a range 
\begin{equation}
H_{min}<H<H_{max}
\label{eq:range}
\end{equation}
We pose the following,
well defined quantitative question: 
\begin{quote}
\textit{What is the maximal number $p_c$ of
different background odorants that our neuron can distinguish from the
target, for any concentration within the prescribed  range }?
\end{quote}
To sharpen the question, we put it in a more precise mathematical form.
Consider $\mu=1,2...p$ 
background odorants with respective concentrations $H^\mu $ in the
range (\ref{eq:range}). 
Odorant $\mu$ is characterized by the $i=1...N$ affinities 
$K_i^\mu $ of the $N$ receptors. According to the RAD model, these affinities
are selected independently from a distribution 
$\psi (K)$ [1]. 

All our numerical results were obtained using for 
$\psi (K)$ the form (note: $K \geq 0$)
\begin{equation}
\tilde{\psi}(K)=\frac K{\sigma ^2}\exp (-\frac{K^2}{2 \sigma ^2}) 
\label{eq:psi}
\end{equation}
The average and variance of this distribution are given by
\begin{equation}
<K>=\sqrt{\frac \pi 2}\sigma   \qquad \qquad 
var\left( \tilde{\psi}(K)\right) =0.65\sigma 
\label{eq:avvar}
\end{equation}
The distributions suggested in \cite{RAD} were Poisson and binomial. With regard 
to the computational limitations of our model, the 
important idea behind the RAD model lies not 
in the exact form of the distribution, 
but on the fact that the affinities can be thought of
as independent random variables. The main computational 
features of our model will not be altered as long as the 
distribution has the following features: it is zero for negative affinities
and it has finite first and second moments. We have 
used $\tilde{\psi}(K)$ since it satisfies the previous 
constraints and is easier to deal with in analytical calculations.

When receptor $i$ is exposed to odorant $\mu$, at concentration $H^\mu $, its
response is given by
\begin{equation}
S_i^\mu=f(K_i^\mu H^\mu )
\label{eq:Si}
\end{equation}
where $f(x)$ is a sigmoid shaped function; we use 
\begin{equation}
f=x/(1+x)
\label{eq:feq}
\end{equation}
throughout this paper. 

The value taken by the
affinity $K_i^\mu $ sets the particular concentration
scale at which odorant $\mu $ 
affects the $i$th sensory neuron. 
From this point on we set
$
\sigma=1
$
in eq. (\ref{eq:psi}); this means
that the concentrations are measured in inverse units of the parameter 
$\sigma .$ 

The set of values $\{ S_i^\mu \}=\{ S_1^\mu,S_2^\mu,...S_N^\mu \}$ 
constitute a vector of signals { \bf S}$^\mu$, 
generated by the entire sensory array, when it is
exposed to odorant $\mu$.
The $\left\{ S_i^\mu \right\} $ serve as inputs to our secondary
neuron, which we model as a linear threshold element or perceptron; its
output signal is given by 
\begin{equation}
 s^\mu =\mathrm{sign} \left( \sum_i^N w_i S_i^\mu \right)  
 =  \mathrm{sign} \left( {\rm \bf w} \cdot {\rm \bf S}^\mu \right)
\label{sinal}
\end{equation}

\begin{figure}
\centerline{\psfig{figure=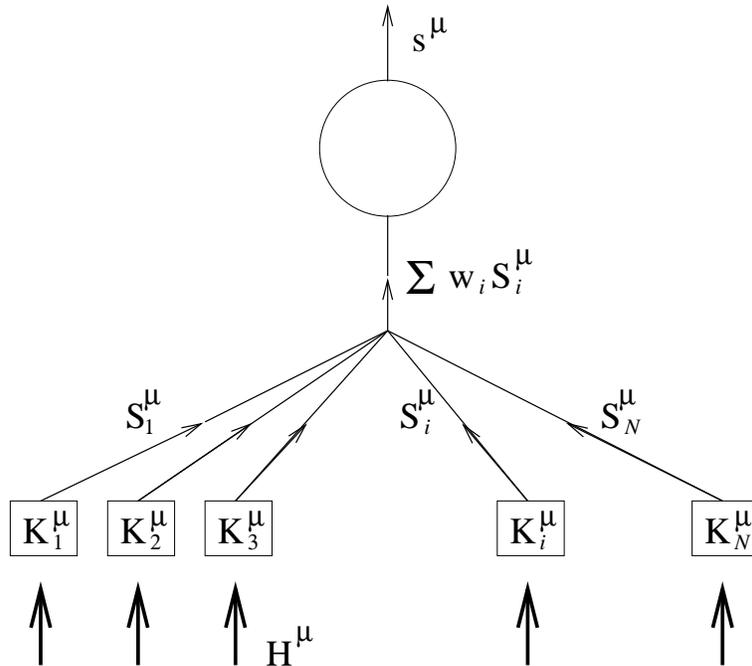,width=10.0cm}
}
\caption{Schematic representation of our model. $N$ sensory neurons,
represented by boxes, are exposed to odorant $\mu$, present in concentration 
$H^\mu$. Each sensory neuron $i$ is characterized by a set of affinities  
$K^\mu_i$; the sensory input elicits from neuron $i$ a response $S_i^\mu$,
as given by eq. (\ref{eq:Si},\ref{eq:feq}). A weighted sum
of the  $N$ sensory responses serves as the input of the secondary neuron
(circle), whose output $s^\mu$, generated in response 
to odorant $\mu$, is given by eq. (\ref{sinal}). 
}
\label{Fig1}
\end{figure}

The simple neural network described above is schematically presented in Fig.
 \ref{Fig1}. The sensory neurons are represented by boxes and the secondary
 neuron by a circle.
We require the output of this neuron to differentiate the target odorant
from the background, i.e. yield

\begin{equation}
s^\mu=
\left\{
\begin{array}{ll}
-1 & \mbox{for  $ \mu=1,...,p$ (background)}  \\ 
+1 & \mbox{for $\mu=0$ (target)}
\end{array}
\right.
\label{eq:smu} 
\end{equation}
for \textbf{any} odorant
concentration in the allowed range (\ref{eq:range}). 

To understand the geometrical meaning of this requirement, note that when 
the concentration of odorant $\mu$ is varied in the allowed range
(\ref{eq:range}), the corresponding vector  {\bf S}$^\mu$ traces a {\it curve}
(or {\it string}) in the $N$-dimensional space of sensory responses. 
The requirement (\ref{eq:smu}) means that there exists a hyperplane, such that 
the entire curve that corresponds to the target odorant lies on one side of it,
while  the curves that correspond to {\it all } $p$ background odorants lie on
the other side. This explains our statement, made in the Introduction, that the
problem we solve deals with the {\it Linear Separability of curves}.

We show that a solution to this
classification problem can be found, provided $p<p_{max}=\alpha _cN$. We
estimate the critical capacity $\alpha _c$ numerically. This is done by 
extrapolating results obtained for various values of $N$, using finite size
scaling techniques, to the limit $N \rightarrow \infty$. The value of $\alpha_c$
is evaluated as a function of the limiting odorant concentrations.

In order to obtain these results using existing methodology, the most natural
and straightforward thing to do is to place a discrete set of 
$\zeta = 1,2,...,M$ points $\mathbf{S}^{\mu \zeta }$ on each
curve, corresponding to different concentrations, 
and to require that the $M$ points that lie on the curve of the target 
odorant
are linearly separable from the $PM$ points that represent the 
background. That is, equations (\ref{eq:smu}) become
\begin{equation}
s^{\mu \zeta} = \left\{ 
\begin{array}{ll}
-1 & \mbox{for $\mu=1,...p; \qquad \zeta=1,...M$ (background) } \\
+1 & \mbox{for $\mu = 0; \qquad \zeta=1,...M$ (target)}
\end{array}
\right.
\label{eq:smuzeta}
\end{equation} 
This raises the technical question of how many
(discrete) representatives  of the same odorant
should be included in the learning set. 
We show below, that while the
critical number of odorants $p_{max}$, scales linearly with $N$, the number
of representatives of a single odorant, $M$, has
to grow at least as fast as $N^2$. This  ensures that increasing 
$M$ further does not change the results of the calculation (e.g.
the value of $p_{max}$) and hence
the $M$ discrete points indeed
represent correctly the continuous curves on which they lie. 

Our problem has been turned into one of learning $M(p+1)$ ``patterns", 
that
constitute our training set $\mathcal{L}$. 
For technical reasons it is
convenient to introduce and work with normalized patterns,
\begin{equation}
\mathbf{\xi }^{\mu \zeta
 }=\frac{{\rm \bf S}^{\mu \zeta }
s^{\mu \zeta }}{\sqrt{\left( {\rm \bf S}^{\mu \zeta }\right) ^2}} 
\label{eq:xi}
\end{equation}
with $\zeta =1.\ldots M$ running over the $M$ discrete concentrations and
$\mu = 0,1,.\ldots p$ over all odorants. Note that we also multiplied each 
pattern {\bf S}$^{\mu \zeta}$ by its desired output, 
$s^{\mu \zeta}$; after this change of
representation the
condition of linear separability (\ref{eq:smuzeta}) becomes
\begin{equation}
\mathrm{sign} \left( {\rm \bf w} \cdot \mathbf{\xi }
^{\mu \zeta} \right) > 0  \qquad \mbox{for $\mu = 0,1,...p$ and
$\zeta=1,2,...M$}
\label{eq:pos}
\end{equation}

\subsection{The Learning algorithm}
The question posed above, 
whether the target odorant can or cannot be distinguished 
from the background, has been reduced to the following one: is there a set of
weights $w_i,~~i=1,...N$, for which all $M(p+1)$ inequalities (\ref{eq:pos})
are satisfied? This problem is of the type studied by
Rosenblatt\cite{Rose}, and is
an example of classification by a single layer
perceptron. A solution exists if one can find a weight
vector $\mathbf{w}^*$ (that parametrizes the perceptron) such that for all the
patterns $\mathbf{\xi }^{\mu \zeta }$ in the training set $\mathcal{L}$
the ``field" 
\begin{equation}
h^{\mu \zeta}=\mathbf{\xi^{\mu \zeta} \cdot w^*} > 0,
\label{eq:field}
\end{equation}
i.e. the projection  of the weight
vector $\mathbf{w^*}$ onto all patterns $\mathbf{\xi}^{\mu \zeta}$
is positive.  We wish to determine
the size of the training set $\mathcal{L}$, i.e. the number of background 
odorants $p$, for which a solution 
$\mathbf{w}^*$ can be found. This is done by executing a search for 
a solution $\mathbf{w}^*$ by means of a {\it learning algorithm}.
There are several learning
algorithms (e.g. Rosenblatt\cite{Rose}, Abbott and Kepler\cite{Abbott} )
in the literature; all are guaranteed to find such a 
weight vector, in a finite number of steps, 
{\it provided a solution exists}. If, however, the problem is {\it not}
\textbf{LS} and  a
solution {\it does not} exist, most learning algorithms will just run ad
infinitum.
An exception to this is the algorithm of Nabutovsky
and Domany (ND)~\cite{ND} which detects, in finite time, 
that a problem is non-learnable. 
This is a batch perceptron learning algorithm,

presenting sequentially the entire training set 
$\mathcal{L}$
in one ``sweep" and repeating the process until either a solution is found or
non-learnability is established. 
We found that this algorithm is efficient and
convenient to use
(see \cite{Abbott} for other algorithms that detect
non-\textbf{LS} problems).
\footnote{The fact that we use a learning procedure to establish the
boundaries of linear separability
doesn't imply and is unrelated to  any possible plasticity
of the olfactory bulb. The algorithm is being used only to either 
find a solution
or show that it doesn't exist.}. 

\begin{figure}
\centerline{\psfig{figure=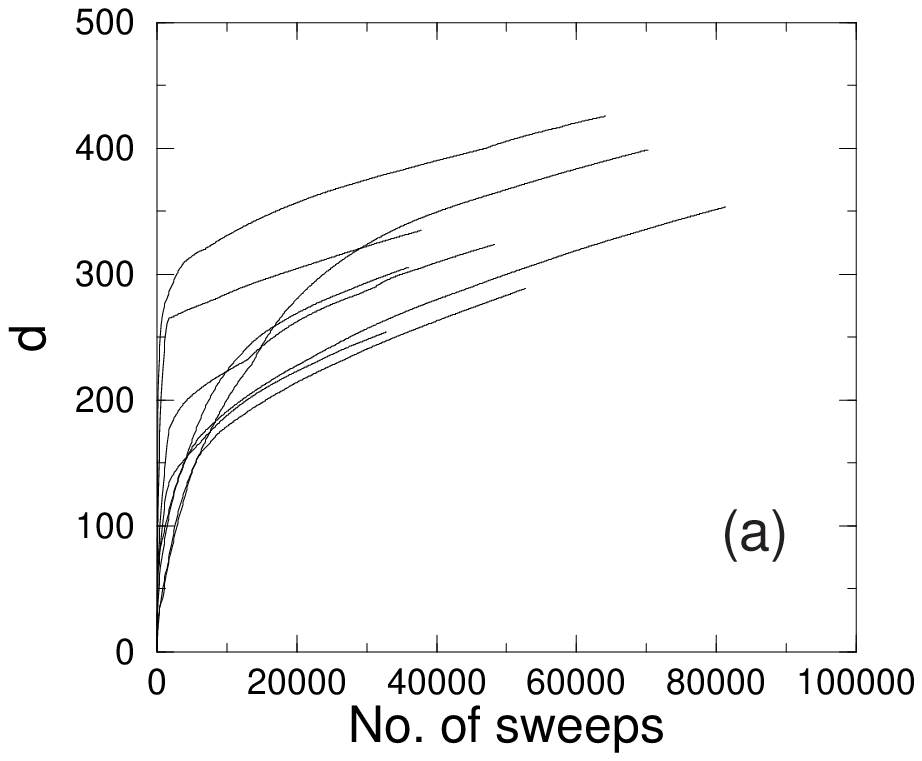,width=7.0cm}
            \psfig{figure=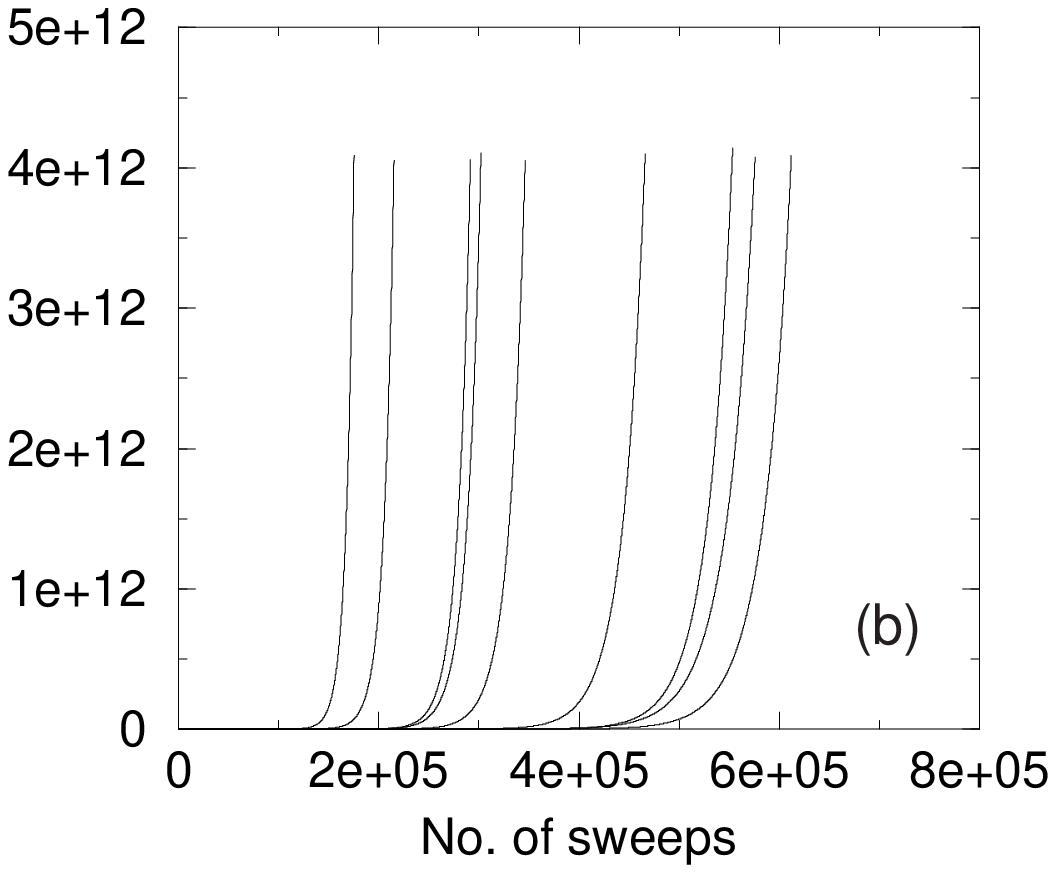,width=7.0cm}
}
\caption{Despair $d$ as a function of the number of sweeps for $N=10, 
H_{max}=50 $, $H_{min}=0.5$ and $p=25$ ;  learnable
cases are shown in (a) while those that were not learnable are shown in (b). Note the
huge difference in both vertical and horizontal scales.
}
\label{fig:dsino}
\end{figure}

ND introduced a parameter $d$ which they called
\textit{despair }, which is calculated ``on line" in the course of the learning
process.  $d$  is bounded if the training set
$\mathcal{L}$ is \textbf{LS}. Since the
ND algorithm can be shown to either find a solution $\mathbf{w^*}$,
or transgress the bound for $d$
in a finite number of learning iterations, 
$d$ effectively signals if the learning
set $\mathcal{L}$ fails to be linearly separable. 
The theorem they proved can be easily
extended to the distribution of examples in our problem
\footnote{%
The original ND algorithm was designed for binary vectors, i.e. pointing at
the corners of a N-dimensional hypercube. It can be shown that the theorem
can be extended to vectors on the unit sphere.}.
We introduced a halting criterion, which is probably more stringent than
necessary, since no attempt has been made
to determine an optimal lower bound. 
In figures \ref{fig:dsino}a and \ref{fig:dsino}b typical evolutions of the despair are shown for
an \textbf{LS} case and for a non-\textbf{LS} case, respectively.
The behavior of $d$ is strikingly different in the two cases, 
showing that indeed $d$ is a good indicator of learnability.
In the learnable cases $d$ grows linearly with the number of learning sweeps
until a solution is found (and the curves terminate). In the non-\textbf{LS}
cases $d$ grows exponentially with the number of sweeps and would continue to
grow; the process is halted when it's value exceeds a known bound, that must be
satisfied if the problem is \textbf{LS}.

We now describe the ND algorithm used in the
simulations. The patterns of the learning set $\mathcal{L}$ are presented
one at a time (one cycle constitutes a sweep). 
ND have shown~\cite{ND} that for binary valued patterns ($\xi_i = \pm1$), i.e.
patterns on
vertices of a unit hypercube, an upper bound $d_c$ exists iff the 
training set is 
\textbf{LS}. 
On the other hand the dynamics is shown to take $d$ beyond that
bound in a finite (linear in $N$) number of iterations unless a solution
exists and the algorithm halts. 
Initialize the process with  $d=1,\,\,\,\,\mathbf{w}=%
\mathbf{\xi }^{1}.$ Go to the next example. If it is correctly classified,
do nothing to the current weight vector and go to the next example. Once a
misclassified example $\mathbf{\xi}^{\mu \zeta}$ 
is found, update  the weight vector
as well as the parameter $d$, according to
\begin{equation}
\mathbf{w}_{new}=\frac{\mathbf{w}+\eta \mathbf{\xi }^{\mu \zeta }}{\left| 
\mathbf{w}+\eta \mathbf{\xi }^{\mu \zeta }\right| }
\end{equation}

\begin{equation}
d_{new}=\frac{d+\eta }{\sqrt{1+2\eta h^{\mu \zeta }+\eta ^2}}
\end{equation}
$\eta $ is not just a learning rate parameter but an effective modulation
function, chosen in order to maximize the increase of the despair as

\begin{equation}
\eta =\frac{1/d-h^{\mu \zeta }}{\left( 1-h^{\mu \zeta }/d\right) }
\label{4.10}
\end{equation}
The learning dynamics halts if all patterns are correctly classified or
alternatively, if the value of $d$ exceeds an upper bound, given by 
\begin{equation}
d>d_c=\frac{N^{\left( N+1\right) /2}}{2^{N-1}}  \label{upperbound}
\end{equation}
This is guaranteed to happen in at most $N$\label{upp}$d_c^2$ steps.
\begin{figure}
\centerline{\psfig{figure=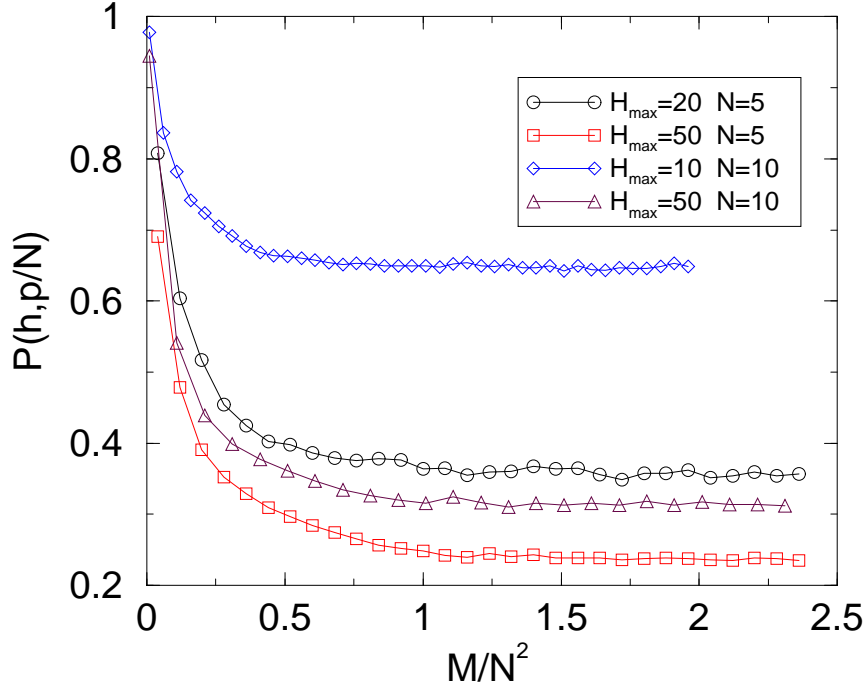}
         }
\caption{The estimated
probabilities saturate when $M$, the number of representatives of a given
odorant, exceeds $M_c \approx N^2.$ For brevity we denote 
$P(H_{min},H_{max},p,N)$ by $P(h,p/N)$.
In all cases $p=25$ odorants were used with $H_{min}=0.5$.}
\label{fig:MNvsP}
\end{figure}

\section{Numerical Experiments}
Since there is a large number of parameters that are to be varied, we 
present first a detailed description of the manner in which we deal with 
every one of
them.

There are two random elements in our studies. The first is in the
selection of $M$ concentrations for each odorant, within the range 
(\ref{eq:range}); the second is the choice of  $K_i^\mu$, the affinity of
receptor $i$ to odorant $\mu$, selected at random from the distribution
$\tilde{\psi}(K)$ of eq. (\ref{eq:psi}). For every choice of the remaining 
variables
we generate an ensemble of experiments and average the object we are measuring
over these two random
elements. We select $L_a$ times the set of affinities and for each of these
perform $L_c$ times the
random selection of concentrations. 

The object we wish to estimate numerically is the probability 
$P$, that the $p$ curves described in the Introduction
are {\bf LS}. To this end we place $M$ points on each curve 
 and measure the corresponding 
probability $P(H_{min},H_{max},p,N;M)$. As we will see, for large
enough values, $M>M_c$, this probability {\it becomes independent of M}; beyond
$M_c$ the set of $M$ discrete points represents the corresponding curves
faithfully and hence the limiting value  $P(H_{min},H_{max},p,N;M >M_c)$
is our estimate for $P(H_{min},H_{max},p,N)$. 

Finally, we are interested in this function in the large $N$ limit, i.e.
when $N \rightarrow \infty$ and $p \rightarrow \infty$, while $\alpha=p/N$
is fixed. This limit is obtained by extrapolating our finite $N$ results,
using finite size scaling methods.

Our first task is to determine how  $M_c$ scales with $N$; that is, how
dense a set of concentrations is to be used so that $M$ discrete points 
represent accurately the continuous curves $S^\mu_i (H^\mu)$ of eq.
(\ref{eq:Si})?

\subsection{Scaling of $M_c$}
We choose values for $N$ (number of receptor cells), $p$ (number of odorants)
and $H_{min},H_{max}$ (limiting concentrations). We also set some value for $M$, the
number of concentrations by which every odorant is represented ($M$ will
will be varied).

We proceeded according to the following steps:
\begin{enumerate}
\item
Draw from the distribution $\tilde{\psi} (K)$
a set of affinities $K_i^\mu$ for all $N$ receptors and 
$p$ odorants. 
\item
Generate for each odorant $M$ concentration values, from a uniform
distribution in the allowed range \newline $H_{min}<H<H_{max}$ and construct 
the set 
$\left\{ \mathbf{\xi }^{\mu \zeta }\right\} $ of normalized patterns.
\item
Run the ND learning algorithm until it stops; register whether the set
$\left\{ \mathbf{\xi }^{\mu \zeta }\right\} $ was {\bf LS} or not.
\end{enumerate}
Steps 2,3 are repeated $L_c$ times for each set of affinities; the whole
process 1 - 3 is repeated for $L_a$ different sets of affinities. 
We used 
$L_a=100$; increasing it further made no difference. With such a value of $L_a$
the results did not depend on $L_c$; having tried $1 \leq L_c \leq 10$
we used $L_c=1$ in our simulations.

At this point we have $L_c \cdot L_a$
experiments, out of which a fraction of $P(H_{min},H_{max},p,N,M)$ cases
were linearly separable. Keeping $H_{min},H_{max},p,N$ fixed, we increase $M$
and repeat the entire process, obtaining the probability functions 
$P(H_{min},H_{max},p,N,M)$, that are plotted in Fig. \ref{fig:MNvsP} 
versus $M/N^2$.
Clearly the curves saturate when $M > M_c \approx N^2$. From  this point 
on we have
fixed the value of $M$ at $M=N^2$. 
This numerical result can be estimated by using  the analysis of 
Gardner and Derrida \cite{GD} for the capacity of biased patterns, 
using for the ``magnetization" $m$ the value $m \propto 1/N$. This gives,
in addition to the leading behavior $M_c \approx N^2$, logarithmic corrections
as well.
We cannot
rule out the
possibility of such logarithmic corrections to the scaling we found here. 

\subsection{Measuring the probabilities $P(H_{max},p,N)$}
In all our experiments we fixed the value of $H_{min}=0.5$ and hence the
dependence of the probability on this variable has been suppressed.
For various values of $N$, $p$ and $H_{max}$ we calculate $P(H_{max},p,N)$
in the manner described above. Keeping $N$ and $H_{max}$ fixed, 
we increase $p$. For $p << N$ we have $P(H_{max},p,N)\approx 1$ and the 
probability of {\bf LS} decreases as $p$ increases. We stop increasing $p$
when $P(H_{max},p,N)$ becomes smaller than some $\varepsilon .$

\begin{figure}
\centerline{\psfig{figure=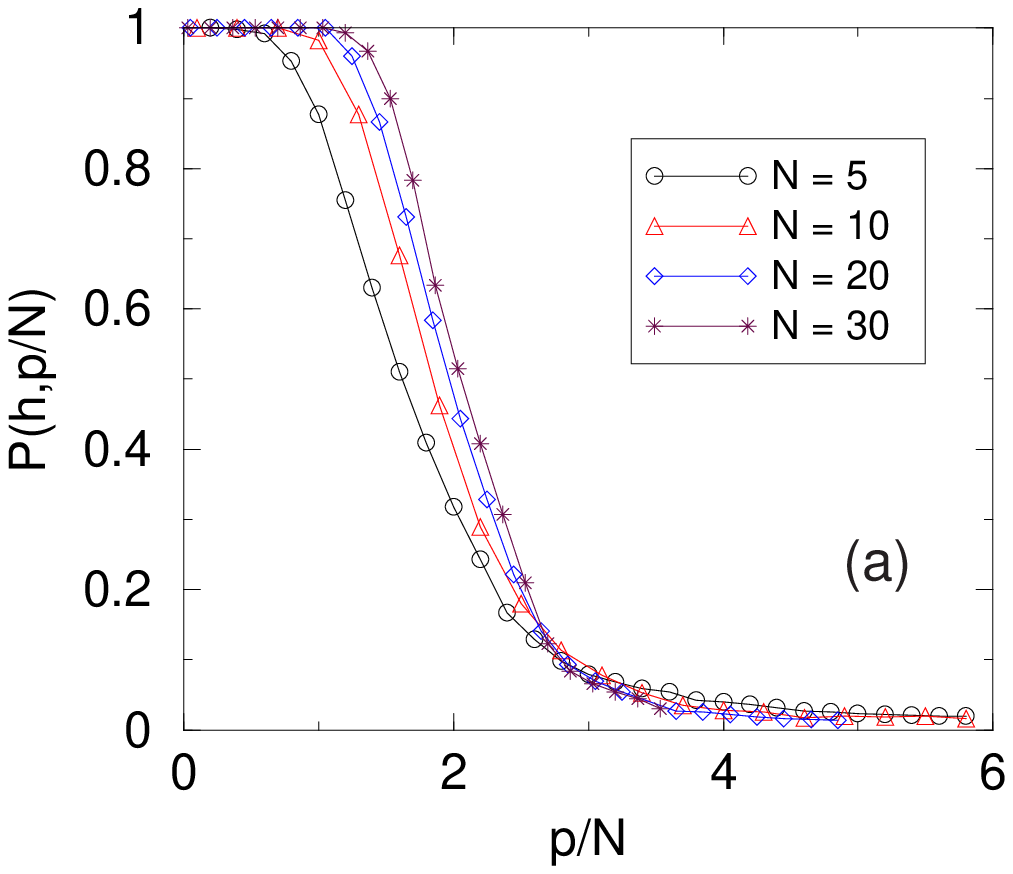,width=7cm}
            \psfig{figure=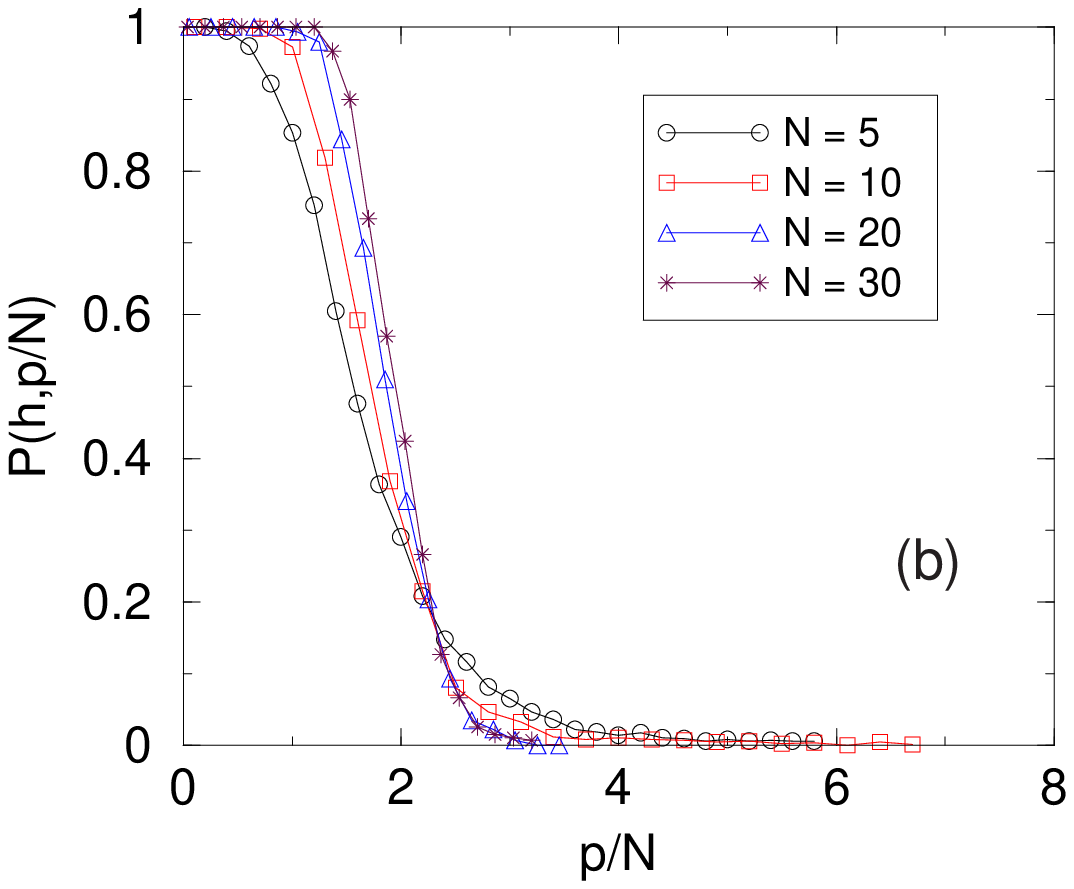,width=7cm}}
\centerline{\psfig{figure=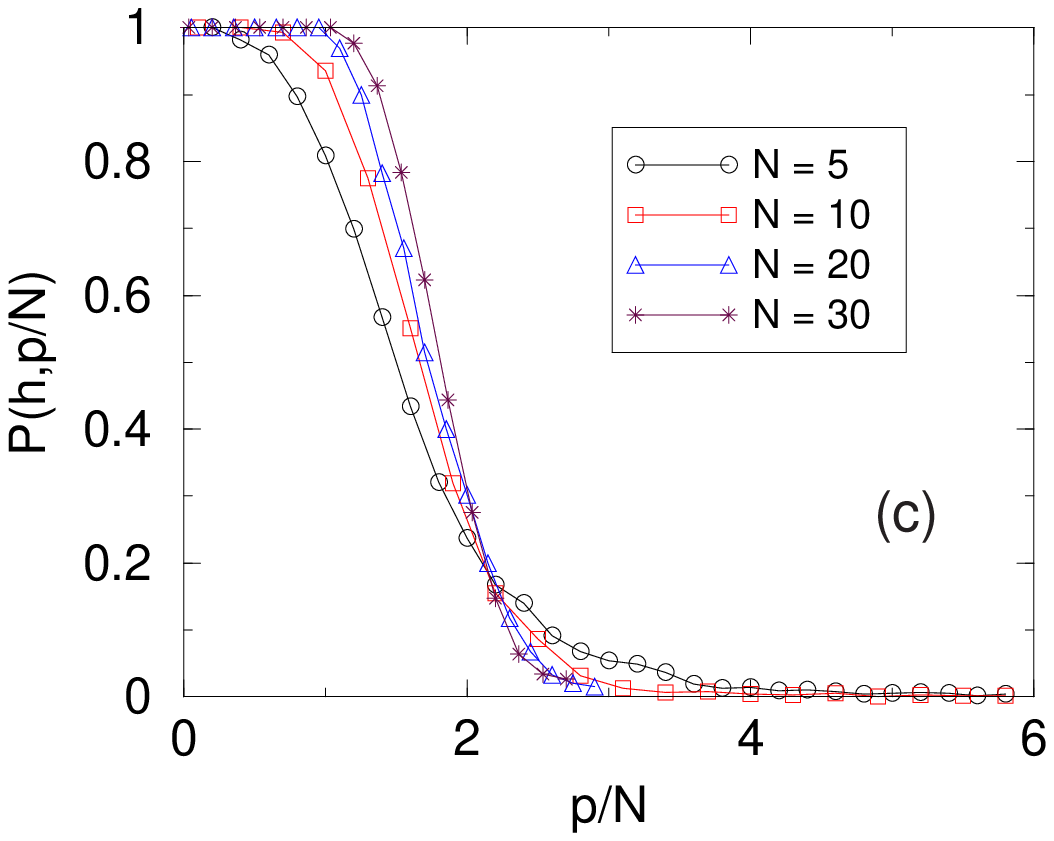,width=7cm}
            }
\caption{
(a) Probability that a
learning set is LS as a function of $\alpha = p/N$, 
for maximal concentration $%
H_{max}=50$ and $H_{min}=0.5$
(b) $H_{max}=70$ and $H_{min}=0.5$
(c) $H_{max}=100$ and $H_{min}=0.5$
}
\label{fig:gfH}
\end{figure}

The variation of  $P(H_{max},p,N)$ vs $p/N$ is presented,  for three values of 
$H_{max}$ and four values of $N$, in Fig. \ref{fig:gfH}. 
The results presented in these
figures are discussed in the next subsection.

We should mention here that for large $N$ we used a heuristic modification of the
ND halting criterion, to label a problem as non-{\bf LS}.
Typical evolutions of the despair parameter are shown in figures \ref{fig:dsino}. Each
curve represents the history for a single learning set. Notice the huge
difference in scales for the learnable and the unlearnable cases. The wide
separation in final values of $d$ suggests that a more practical, e.g.
smaller, upper bound be used. For $N=30$ (the largest value treated here)
we used a different halting criterion in order to escape from the need to
reach an exponentialy high upper bound. After a small number of successful 
trial runs (that did produce linear separabitiy) we identified 
the highest     value of the
despair $d_m$ that was reached for a learnable set. This value was used to
define our new heuristic halting criterion, $d_{bound}=N$ $d_m$.

\subsection{Finite Size Scaling Analysis}
As expected, for small $\alpha = p/N$ the probability for linear separability 
is close to 1, and it decreases as $\alpha$ increases. The curves obtained for
fixed $H_{max}$ become sharper as $N$ increases.

%
%
%
Note that 
curves obtained for different $N$ values 
cross at approximately the same value of $\alpha$. Similar behavior
of the corresponding probability functions has been observed for
random uncorrelated patterns~\cite{Cover}. 
Notice, however, that the crossing point is at 
some probability $P < 1/2$. Similar curves, obtained for 
other architectures, such as the
parity and commitee machines \cite{nadler} crossed at $P>1/2$. If there is
a sharp transition in the thermodynamic limit ($N \rightarrow \infty$),
these curves should approach a step-function, with 
\begin{equation}
P(H_{min},H_{max},\alpha)=\left\{ \begin{array}{ll}
1   &   \mbox{if $\alpha < \alpha_c(H_{min},H_{max})$} \\
0   &   \mbox{otherwise}   \end{array} \right.
\label{eq:alphac}
\end{equation}
That is, for $\alpha $ below a certain $\alpha _c\left(
H_{min},H_{max}\right) ,$ a learning set will be \textbf{LS} with
probability one and conversely, it will be \textbf{LS} with probability zero
for $\alpha >$ $\alpha _c\left( H_{min},H_{max}\right) .$ 
The manner in which such a step function is approached as $N \rightarrow \infty$
can be described by a finite size scaling analysis (e.g. \cite{FSS}). 

\begin{figure}
\centerline{\psfig{figure=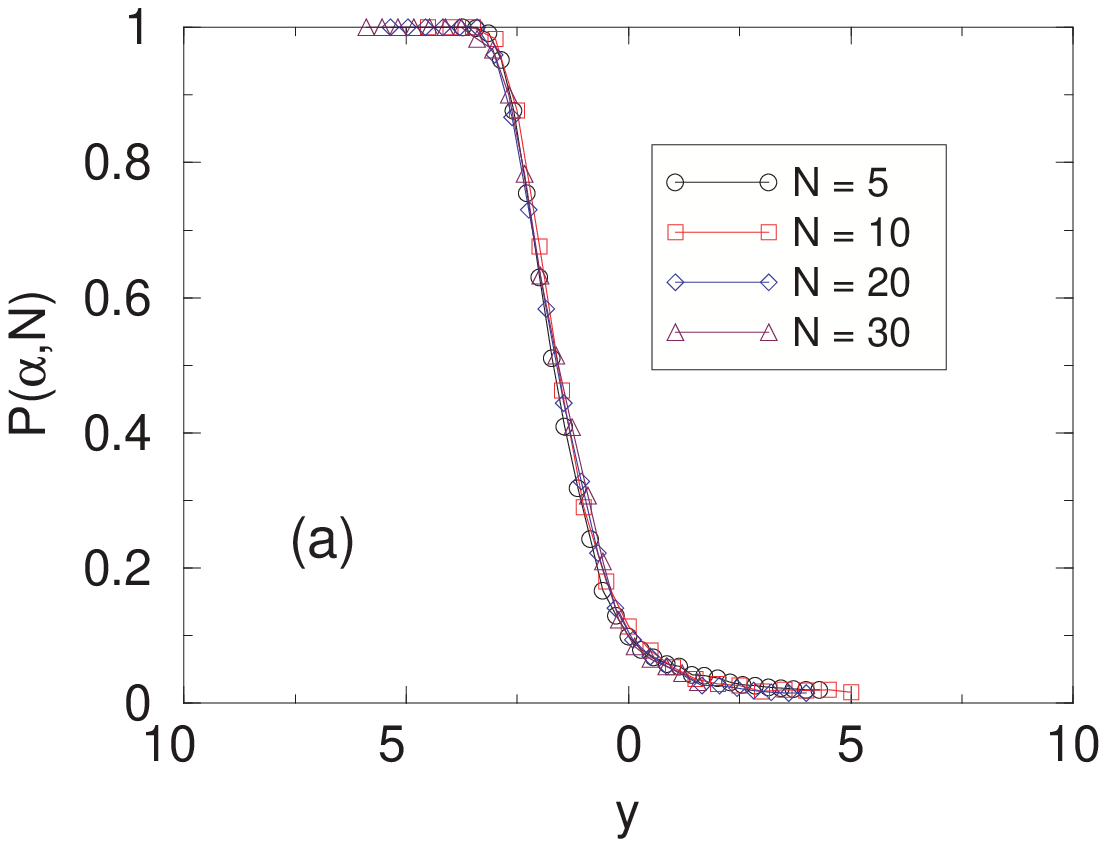,width=7cm}
\psfig{figure=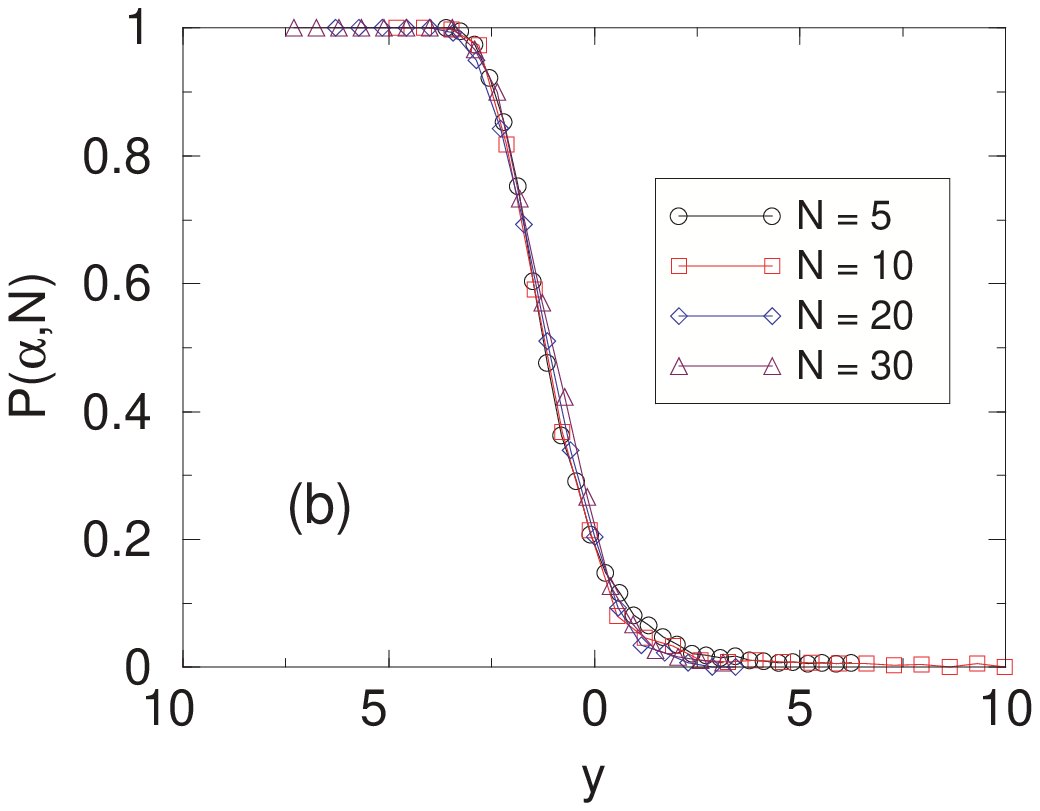,width=7cm}}
\centerline{\psfig{figure=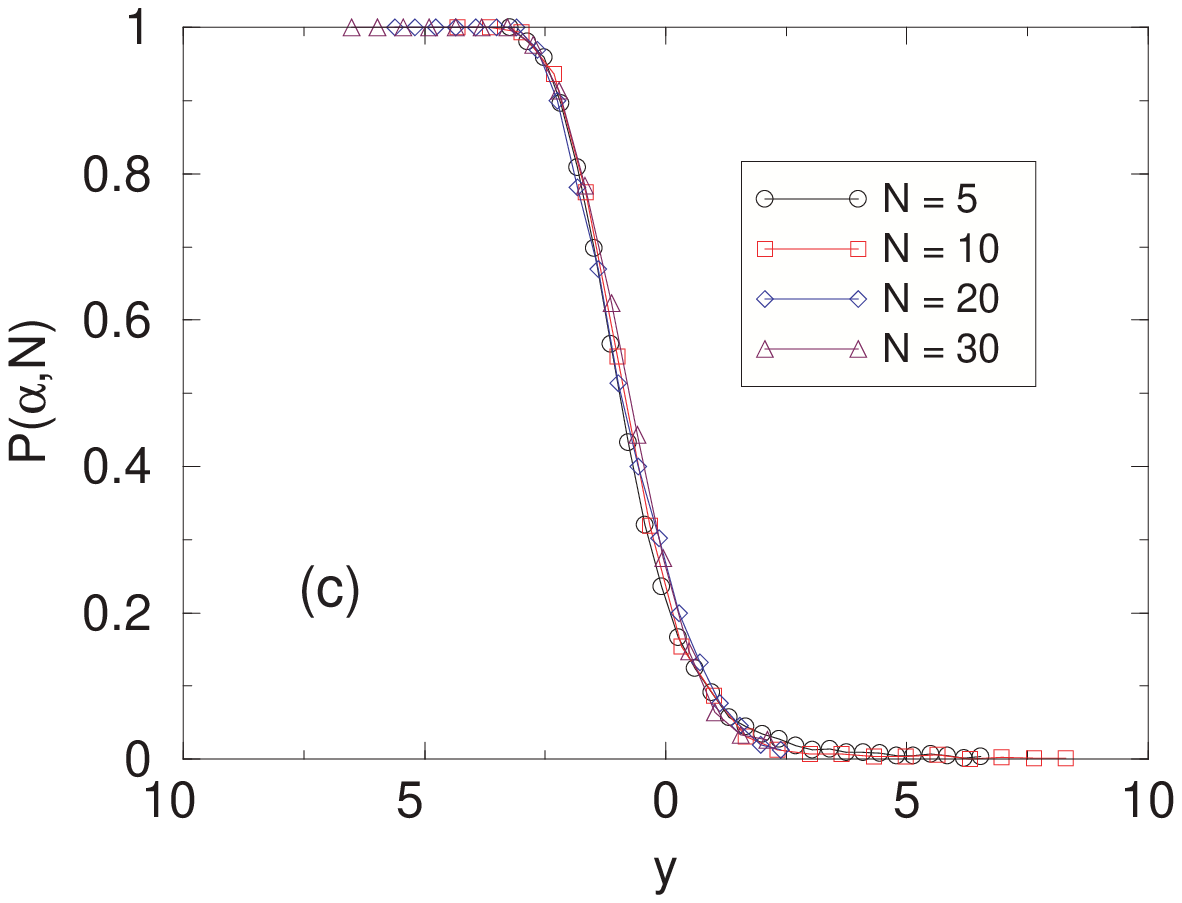,width=7cm}
            }
\caption{
(a) Probability that a
learning set is LS as a function of $y=(\alpha -\alpha _c)N^{\frac 1\nu }$, 
for maximum concentration 
$H_{max}=50$ and $H_{min}=0.5$
(b) $H_{max}=70$ and $H_{min}=0.5$
(c) $H_{max}=100$ and $H_{min}=0.5$
The result of a least squares fit are
(a)$H_{max}=50,\nu =4.8,\alpha _c=2.8,$(b) $H_{max}=70,\nu =2.85,\alpha
_c=2.25$ (c) $H_{max}=100,\nu =2.75,\alpha _c=2.1$
}
\label{fig:scalH}
\end{figure}

For each value of $H_{max}$ (keeping $H_{min}$ fixed) we tried 
a simple rescaling of the $\alpha $ variable, with two adjustable parameters,
$\alpha_c$ and $\nu$;
\[
y=(\alpha -\alpha _c)N^{\frac 1\nu }. 
\]
For the proper choice of $\alpha_c$ and $\nu$ we expect {\it data collapse}; that
is, curves obtained for different values of $N$ are expected to fall onto a
single function, provided $P(H_{max},\alpha ,N)$ is plotted versus the scaled
variable $y$. As can be seen on
Figures \ref{fig:scalH}a,b and c, this expectation is borne out; 
the evidently good data colapse indeed
substantiates the idea of a sharp transition at $\alpha _c$. 
As $N$ increases, the function $P(H_{max},\alpha ,N)$ becomes increasingly sharper;
its width near $\alpha_c$ decreases at a rate governed by the exponent $\nu$.

%
%
%
%

\begin{figure}
\centerline{\psfig{figure=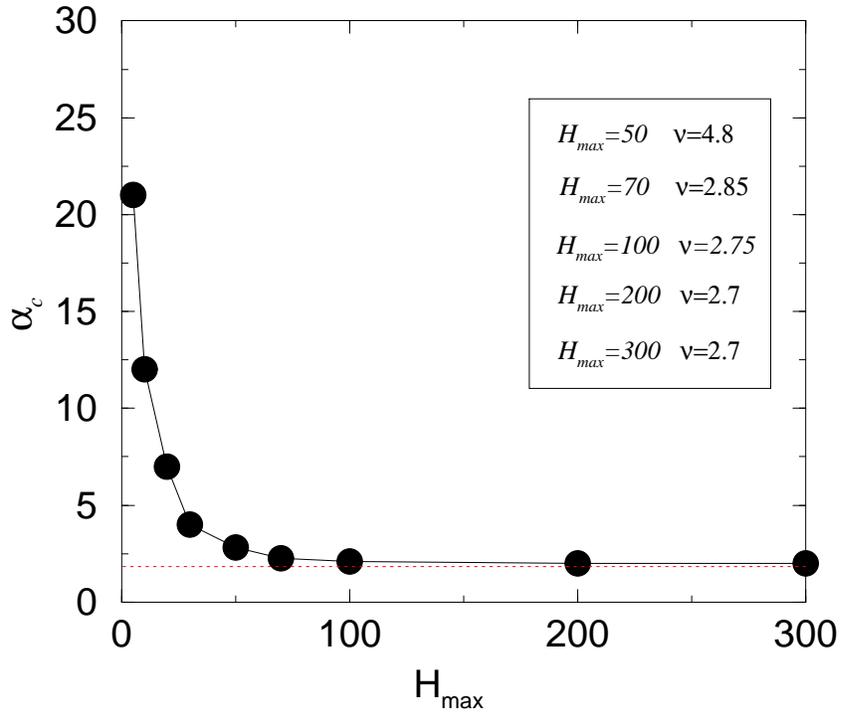}}
\caption{Phase diagram in the 
$\alpha ,H_{max}$ plane for $H_{min}=0.5$.The curve separates the $\alpha
,H_{max}$ plane into the regions where the network can distinguish the
target odorant (below), and where it cannot (above). Below(Above) the curve
a learning set is(not) \textbf{LS}  with probability one in the
thermodynamic limit. The horizontal dotted line is $alpha_c=2.$}
\label{fig:alphac}
\end{figure}

Finally, we present 
in figure \ref{fig:alphac} the behavior of  $\alpha _c$ as a function of $%
H_{max}$ (for fixed $H_{min}$). As $H_{max}$ increases, separation of the curves
becomes an increasingly difficult task and hence $\alpha _c (H_{max})$ 
decreases. We find that 
it saturates at a low value close to $\alpha _{\min }=2$, which is
exactly the Cover result. This interesting point is explained in the Appendix.

Note that even though we deal here with linear separability of {\it curves}, which
one would expect to be a more difficult task than separating points, we found that
our $\alpha_c$ exceeds the value derived for points, $\alpha_c = 2$. The reason is 
that this is the critical capacity for separating {\it random, independent} points;
the curves we are trying to separate are {\it not independent of each other}. In fact
by construction we have $S_i^\mu > 0$ for all the background odorants; hence all 
these curves  lie on one side of an entire family of planes. The target odorant,
which also satisfies $S_i>0$,
should lie on the other side of the separating plane.  

The curve $\alpha _c (H_{max}) $ is, in effect, a {\it phase boundary}; on one side
we have
a ``phase" in which the problem is {\bf LS}, while on the other (high $\alpha$) region
it is not.      We present now a brief description of the manner in which
linear separability breaks down as we cross this phase boundary by
increasing $H_{max}$ at fixed $\alpha$.

\subsection{Breakdown of {\bf LS} near phase boundary}
\begin{figure}
\centerline{\psfig{figure=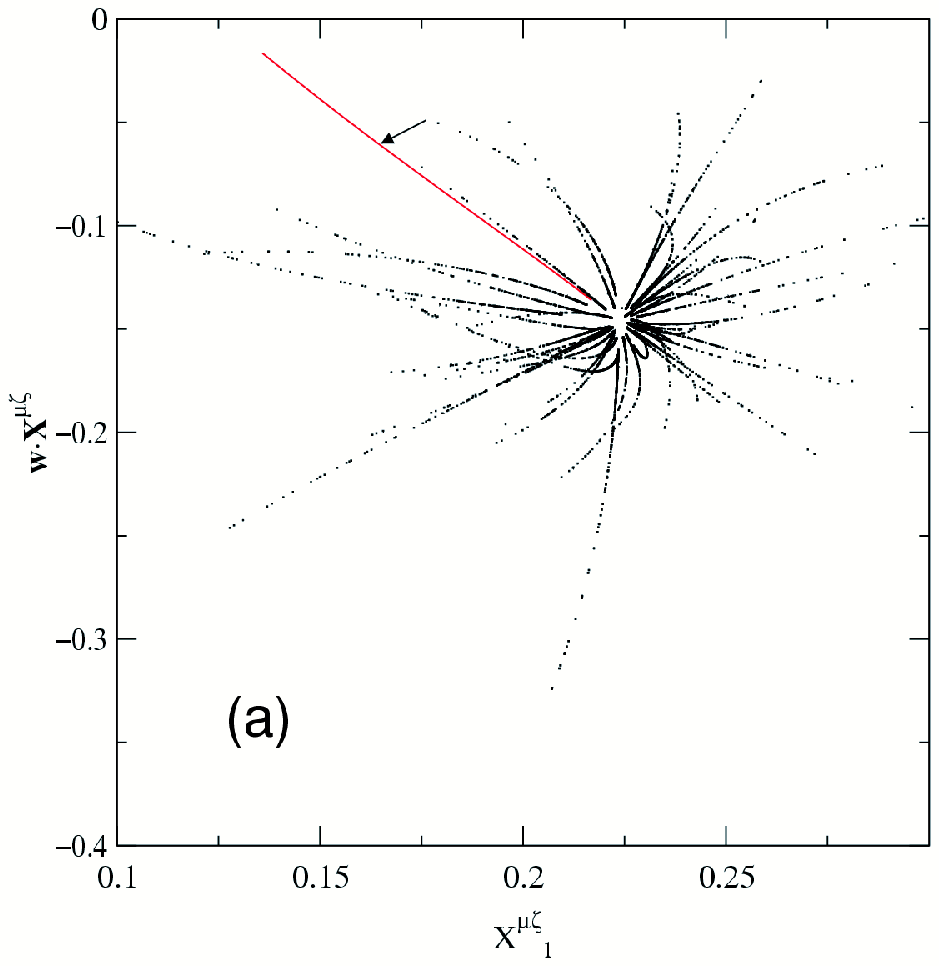,width=7cm}
            \psfig{figure=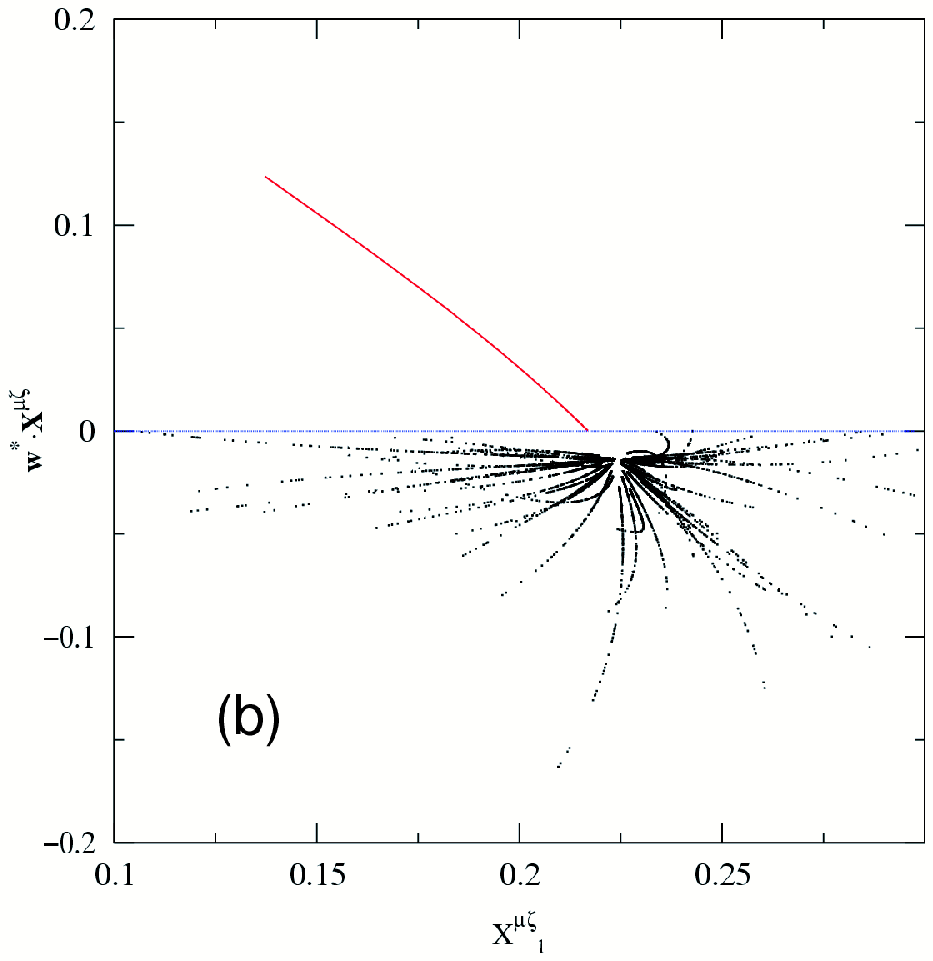,width=7cm}}
\caption{The curves representing the $p$ background odorants and
the target odorant (marked by an arrow), in a {\bf LS} case,
$\alpha <\alpha _c$, for $H_{max}=50,H_{min}=0.5,N=20.$
In figure (a) the curves are projected onto a randomly selected plane; 
in (b) onto a plane $\cal Q$ (see text), that contains the
weight vector $\mathbf{w}^{*}$, determined by the learning algorithm. 
The horizontal dotted line in figure (b)
demonstrates linear separability of the target from background.
}
\label{fig:LS}
\end{figure}

\begin{figure}
\centerline{\psfig{figure=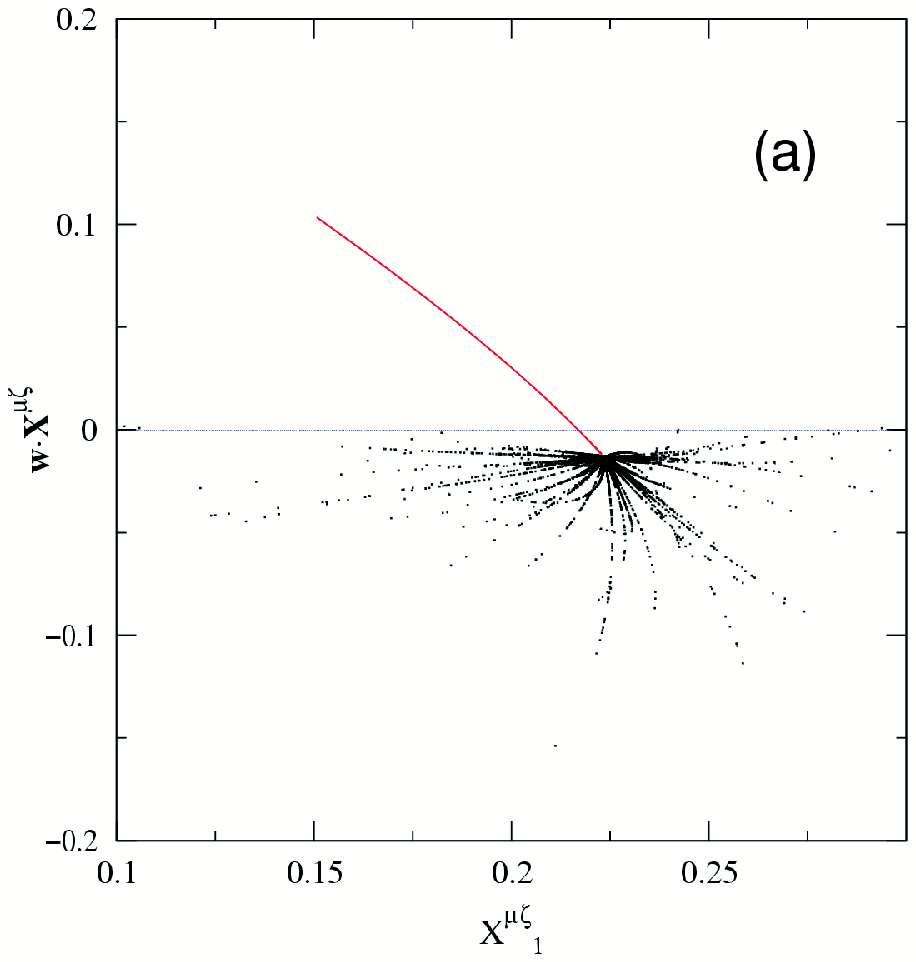,width=7cm}
            \psfig{figure=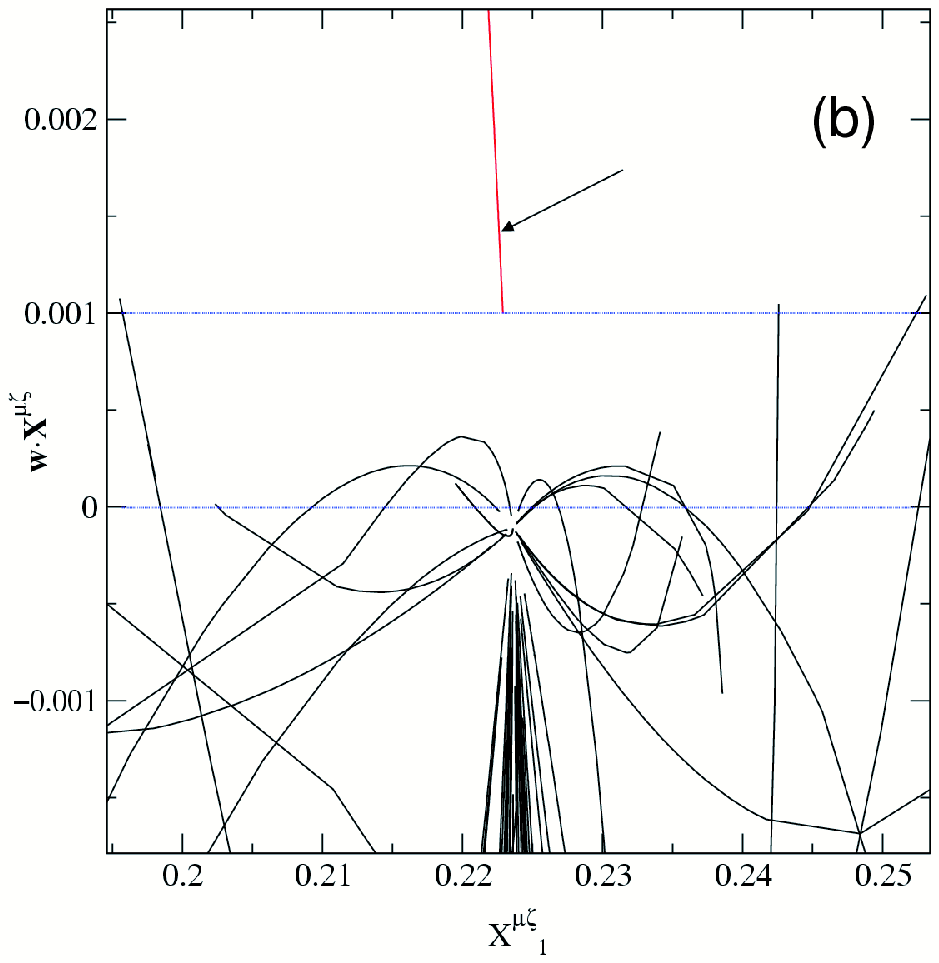,width=7cm}}        
\caption{The same situation as in 
Fig. 7, but with 
$H_{max}$
increased beyond the limit of learnability. (a) Projecting the curves onto 
the same plane as for the
{\bf LS} case, we see that the target penetrates to the "wrong" side of the
broken line. (b) Further attempts
to learn new {\bf w} will fail.
}
\label{fig:NLS} 
\end{figure}

The manner in which {\bf LS} breaks down as $H_{max}$ increases beyond
the phase boundary is nicely illustrated by the set of figures 
\ref{fig:LS} and \ref{fig:NLS}. 

Consider the $p$
curves, in an $N$-dimensional space, which represent the odorants, in 
a linearly separable case. We present in figure \ref{fig:LS}(a) a projection 
of these curves 
onto a randomly chosen plane. One of these (indicated by an arrow)
is the target odorant; it seems to be entangled with the other curves.
The point at which all curves seem to converge corresponds to the maximal 
concentration.

The purpose of
the learning dynamics is to find a particular direction {\bf w}, along  
which one is able to
separate the target curve from the others. 
Denote by $\cal V$ the hyperplane that passes through the origin and is
{\it perpendicular} to {\bf w}; this 
is the linear manifold that separates the target from all the background curves.
Select now any plane $\cal Q$,
that
contains {\bf w}, and project all curves onto $\cal Q$; this
produces Fig. \ref{fig:LS}(b). 
The horizontal dotted line shown here is the intersection
of the hyperplane $\cal V$ with the plane $\cal Q$.
The projected background odorant curves lie on one side of this line and the
target on the other. The situation depicted here is {\bf LS}.

Consider now what happens when we turn the problem into non - {\bf LS}  by 
increasing $H_{max}$ beyond the phase boundary. 
As we increase the maximal concentration, 
the target odorant's
curve penetrates to the ``wrong" side of the hyperplane $\cal V$. 
A picture of this situation is shown in  Fig. \ref{fig:NLS}(a).
This is a non {\bf LS} problem - which means that no matter how long we run our
learning algorithm, we will never find a hyperplane $\cal V$  that separates the
target from all the background. If nevertheless
we keep running our learning algorithm, the direction of our candidate
for {\bf w} will keep changing as we ``learn", but
since the critical capacity curve of figure \ref{fig:alphac} 
has been crossed, no amount
of further learning will produce a separating plane. The density of
points near the high concentration limit is much larger than for low
concentrations. Hence further learning will perhaps be able to
separate the target from the background at high concentrations - but
then separability breaks down at low concentrations (see Fig. \ref{fig:NLS}(b)).

\section{Summary and Discussion}
In the olfactory bulb of most vertebrates, each secondary neuron (mitral  
or tufted cell) 
receives input from only one glomerulus, which in turn is innervated, in 
all likelihood, 
by axons stemming from olfactory epithelial sensory cells that all express 
the same olfactory 
receptor protein. Thus, the grandmother cell modeled here may not 
simply represent a mitral or tufted cell. However, when the network of 
periglomerular and granule cells (interneurons) 
is taken into account, then it is fair to state that each mitral cell 
receives (indirect) input from a 
large number of different olfactory 
receptor types. Thus, the present analysis may be 
relevant to the kind
of neuronal processing that takes place in the first neuronal relay 
station of the olfactory pathway,
the olfactory bulb. Alternatively, it may represent, in abstract fashion, 
information processing that
takes place both in the olfactory bulb and at higher olfactory central 
nervous system centers.

Previously, several studies have been published that analyze neuronal 
networks for the olfactory
System \cite{Free} \cite{Hop}\cite{ZhaoHop} \cite{WB} \cite{Zhao} \cite{ZhaoHertz}. 
However, none of these was based on 
a quantitative model for the affinity relationships within the entire olfactory 
receptor repertoire. Here, we use the Receptor Affinity Distribution (RAD) model, 
which was developed, based on general biochemical considerations, for 
receptor repertoires, including that of olfactory receptors. 
The power of this approach 
is in utilizing a global knowledge about the repertoire to analyze the 
fidelity of discrimination among odorants. It has been pointed out in the 
past, that the RAD model may be used to analyze the signal to noise ratio 
in systems in which specific binding to a receptor has to be distinguished 
from the background of numerous other receptors which constitute 
"non-specific binding"  \cite{RAD} \cite{HOROWITZ LANCET KATZIR 94}. 
Here, we apply a similar concept to an analysis of signal to noise 
discrimination in the case of a neuronal network whose input stems 
from a receptor repertoire.

The results presented here suggest that for a fixed number of background odorants
there is a maximal odorant concentration 
beyond which odorant discrimination becomes impossible. This is not surprising, 
since olfactory receptors are saturable, and at very high concentrations weak affinity 
receptors as well as high affinity ones will generate comparable signals. 
However, it is noteworthy that despite the fact that information capacity for 
odorant discrimination rapidly declines as odorant concentration goes up, the 
presently analyzed network is still capable of discrimination even at concentrations
for which $H \langle K \rangle$ is of the order of a few hundred 
(where $\langle K \rangle$ is the average affinity).

The model network consists of $N$ sensory neurons, each of which
is characterized by a set of affinities to a number of odorants. When 
any particular odorant, $\mu$, is present, sensory neuron $i$ produces a 
(nonlinear) response,  $S_i^\mu$. These responses constitute the inputs to
a single processing unit (secondary neuron), which performs weighted summation
of all the $N$ inputs. The secondary neuron's output is the sign of this
weighted sum. The aim of this single processing unit is to identify {\it one
single odorant} separate it from all the others that may be sensed by the
system. This secondary neuron plays the role of a "grandmother cell" for a 
particular target odorant.
An assemply of $P_0$ such secondary neurons may constitute, 
together with the sensory neurons, a system that is able to clearly  identify
the presence of $P_0$ target odorants, from a background of $P$ odorants.

We posed a well defined quantitative question: given that each odorant may
appear with a concentration $H_\mu$ that lies in a certain range, 
$H_{min} < H^\mu < H_{max}$, what is the maximal
number of background odorants $P_c=\alpha_c N$, from which a 
single target can be separated
with probability 1? The answer is summarized in Fig. 5, where $\alpha_c$, the
critical capacity, is plotted vs. $H_{max}$. The result is obtained in the limit
of large $N$ (i.e. many sensory neurons - in fact, for $N=100$ 
this result should
already give excellent precision). For a dynamic range of $H_{max}/H_{min}$ of
about 100 we find $\alpha_c \approx  2.5$. That is, for say $N=300$ sensory
neurons we can distinguish the target from about 750 background odorants.
Hence if we assemble 750 odorants and appoint a grandmother cell for each, we
will be able to identify them one by one.

In order to get this quantitative answer we had to generalize an old problem,
of {\it Linear Separability} of $P$ points on an $N-1$ dimensional hypersphere,
to the new problem of linerly separating $P$ {\it curves} that lie on the 
same hypershere. We have shown that in order to represent a curve by discrete
points that lie on it, we have to place $M \propto N^2$ points on each curve.
The results were obtained by a perceptron learning algorithm that signals when a
problem is {\it unlearnable}, i.e. non-linearly-separable.

Results obtained at various values of $N$ were shown to collapse when plotted as
functions of properly defined scaled variables, which allowed easy extrapolation
to large values of $N$.

\vspace{1cm}
\centerline{{\bf APPENDIX}}
\vspace{1cm}

The behavior of the phase boundary for large concentrations $\alpha
_c\rightarrow 2$ (figure 5) is quite surprising since the network may be
expected to enter a totally confused state due to the saturation of the
nonlinear sensory neurons. This could be expected to lead instead to $\alpha
_c\rightarrow 0.\;$That the Cover result $\left( \alpha _c=2\right) $ is
recovered in the high concentration regime can be in fact be understood by
the following argument.

We first calculate the probability $P(S),$ that a sensory unit gives a
response $S$ to the presentation of an odorant in the range (1), by
\begin{equation}
P(S)=\langle \delta \left( S-f(HK)\right) \rangle   
\end{equation}
where the average is taken over possible concentrations $H$ uniformly
distributed in range (1) and according to the RAD model, over the
affinities,  $\tilde{\psi}(K)$ of equation (2). $f$ is given by equation
(6). The integrals lead to 
\begin{equation}
P(S)=\frac{\sqrt{\pi }\left( 1-S\right) ^{-2}}{\sigma \left(
H_{max}-H_{min}\right) }\left( Erfc\left( \frac S{\sigma H_{max}\left(
1-S\right) }\right) -Erfc\left( \frac S{\sigma H_{min}\left( 1-S\right)
}\right) \right) ,  
\end{equation}
where $Erfc\left( x\right) =\int_0^x\exp \left( -u^{2}\right) du/\sqrt{\pi }
$ is the complementary error function This probability has one peak which
sharpens and moves to higher values of $S$ as $H_{max}$ grows. However at
the very ends of the interval, $S=1$or $0,$ the probability is zero. That $%
P(1)=0$ for every $H_{max}$ is the source of the surprise. The peak which
concentrates all the probability , gets arbitrarily close to $S=1$, as the
concentration increases, but never makes it to the extreme of the interval. In fact
$S_{peak} \approx 1-c/H_{max}$.
Therefore the components of the vectors $\mathbf{S}^{\mu \zeta }$ will be
with overwhelming probability at the peak position, which can be written as $%
S_i^\mu =1-\varepsilon _i^\mu $ with all $\varepsilon _i^\mu
(H_{max},K_i^\mu )\ $small but strictly positive. Neglecting second order
terms in $\varepsilon $ the normalized patterns will then be:
\begin{eqnarray*}
\tilde{S}_i^\mu  &\equiv &\frac{S_i^\mu }{\left| \mathbf{S}^\mu \right| }=%
\frac{\left( 1-\varepsilon _i^\mu \right) }{\sqrt{N}\left( 1-2\bar{%
\varepsilon}^\mu \right) ^{\frac 12}} \\
&=&\left( 1-\varepsilon _i^\mu +\bar{\varepsilon}^\mu \right) /\sqrt{N}
\end{eqnarray*}
Therefore the $\mathbf{\tilde{S}}^{\mu \zeta }$ vectors are unbiasedly
distributed around $\left( 1,1,\ldots 1\right) /\sqrt{N}$. We are taken back
to the original Cover-Gardner problem of separating $p$ unbiased patterns
with a hyperplane and the result $\alpha _c=2$ is no longer a surprise. This argument
doesn't deal with the asymptotic behavior of the capacity in the presence of any kind of
noise. In that case the naive expectations that $\alpha
_c\rightarrow 0$ for $H_{max}\rightarrow \infty $ are probably borne out.

{\bf Acknowledgements}
We thank Ido Kanter for most useful
discussions.
The research of ED was supported by the Germany-Israel Science Foundation (GIF), the
Minerva Foundation and the US-Israel Binational Science Foundation (BSF).
The work reported here was initiated during visits of NC to the Weizmann
Institute, that were supported by grants from the S\~ao Paulo Society of Friends of the
Weizmann Institute and by the  Gorodesky Foundation. JEPT's 
research was supported by a 
graduate fellowship of the 
Funda\c{c}\~ao de Amparo \`a Pesquisa do Estado de S\~ao Paulo
(Fapesp). NC received partial support
from the Conselho Nacional de 
Desenvolvimento Cient\'{\i}fico e Tecnol\'ogico (CNPq)

\end{document}